\documentclass{elsart} 
\usepackage{epsfig} 
\usepackage{amstext}
\begin{document}
\bibliographystyle{elsevier}
\begin{frontmatter}
  \title{A Compton Backscattering Polarimeter for Measuring
    Longitudinal Electron Polarization}
  \author[NIKHEF]{I. Passchier\thanksref{address}}, %
  \author[Virginia]{D. W. Higinbotham}, %
  \author[NIKHEF]{C. W. de Jager\thanksref{TJNAF}}, %
  \author[Virginia]{B. E. Norum}, %
  \author[NIKHEF]{N. H. Papadakis\thanksref{Greece}} and %
  \author[NIKHEF]{N. P. Vodinas\thanksref{Greece}} %
 
  \address[NIKHEF]{National Institute for Nuclear Physics and High
    Energy Physics, P.O.Box 41882, 1009 DB, Amsterdam,
    the Netherlands} %
  \address[Virginia]{Department of Physics, University
    of Virginia, Charlottesville, VA 22901, USA} %
  \thanks[TJNAF]{present address: Thomas Jefferson National
    Accelerator Facility,12000 Jefferson Avenue, Newport News, 
    VA 23606, USA} %
  \thanks[Greece]{present address: Institute of Accelerating Systems and
    Applications, P.O.Box 17214, 10024 Athens, Greece}
	\thanks[address]{ Corresponding author; tel +31 20 592 2147; 
	fax +31 20 592 5155; email: igorp@nikhef.nl }
\begin{abstract}
Compton backscattering polarimetry provides a fast and accurate
method to measure the polarization of an electron beam in a
storage ring.  Since the method is non-destructive, the polarization
of the electron beam can be monitored during internal target
experiments.  For this reason, a Compton polarimeter has been
constructed at NIKHEF to measure the polarization of the
longitudinally polarized electrons which can be stored in the AmPS
ring.  The design and results of the polarimeter, the first Compton
polarimeter to measure the polarization of a stored longitudinally
polarized electron beam directly, are presented in this paper.
\end{abstract}


\end{frontmatter}

\section{Introduction}
 
Stored polarized electron beams are now available at many
laboratories. Such beams are produced either by radiative
polarization\cite{cbp:bar93,cbp:knu91} of the stored beam or by
injecting polarized electrons.  Internal target experiments performed
with such a beam require the electron polarization to be measured
during the experiments.  These measurements need to be done quickly,
so that any changes in polarization can be accounted for, and
parasitically, so that the measurement in no way affects the internal
target experiment.  One technique to do this is through the use of a
Compton backscattering polarimeter.
 
In this technique, a circularly polarized photon beam is scattered off
a polarized electron beam.  Due to the asymmetry in the spin-dependent
Compton scattering cross section, the polarization of the electron
beam can be determined by changing the helicity of the photon beam.
For transversely polarized electrons the asymmetry is measured with
respect to the orbital plane of the electrons, while for
longitudinally polarized electrons the asymmetry is measured in the
energy-dependent cross section.

\begin{figure}
\epsfig{figure=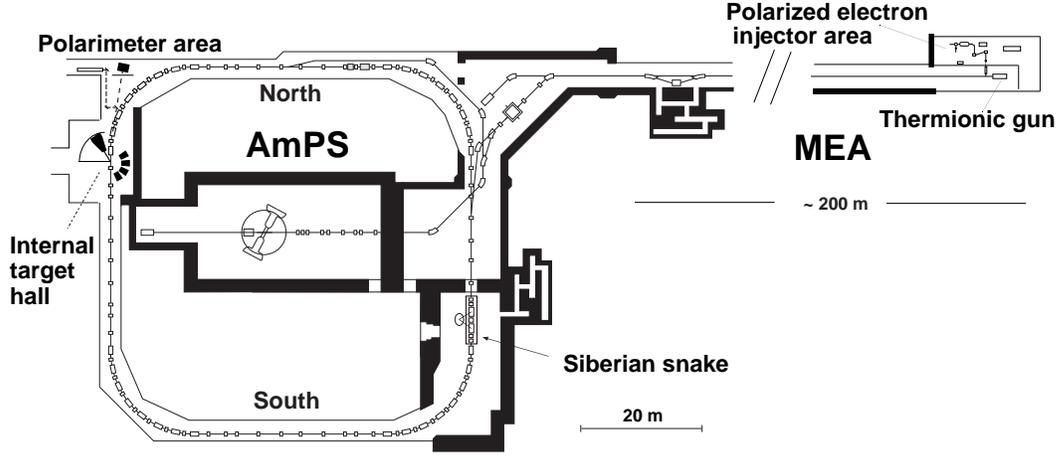,width=\textwidth}
\vspace{4mm}
\caption{The layout of the MEA/AmPS electron facility the equipment 
  for the polarized electron beam.}
\label{cbpfig:amps}
\end{figure}

To perform spin-dependent electron scattering experiments, the
MEA/AmPS facility at NIKHEF has been upgraded to provide a
longitudinally polarized electron beam for internal target
experiments\cite{cbp:bol96}. An overview of the NIKHEF electron beam
facility is shown in fig.~\ref{cbpfig:amps}.  Polarized electrons are
produced by a recently commissioned Polarized Electron Source (PES)
\cite{cbp:bol96}, consisting of a laser driven photocathode electron
source, a Z-shaped spin manipulator to orient the electron spin in any
arbitrary direction, a Mott polarimeter to measure the electron
polarization at the source, and a post-accelerator to match the energy
of the polarized electrons to the acceptance of the chopper/buncher
system of the linac(400~keV). After the post-accelerator, the
polarized electrons are injected into the linear medium-energy
accelerator (MEA) and accelerated up to 750 MeV.  The electrons are
then injected into the AmPS storage ring.  Due to the low current
produced by the polarized source, only a few mA can be stored per
injection, but through the use of stacking a current of more than
200~mA has been stored in the ring.  A Siberian Snake
\cite{cbp:der73,cbp:der75} has been installed in the ring to preserve
the polarization of the stored beam and to maintain a longitudinal
orientation of the spin at the location of the internal
target\cite{jps97}.
 
In this paper the performance of a laser backscattering polarimeter is
presented, which was designed and constructed for the measurement of
the longitudinal polarization of the stored electron beam in the AmPS
ring.  While many laboratories use Compton backscattering polarimeters
to measure the polarization of stored transversely polarized electron
beams\cite{cbp:bar93,cbp:bar94,cbp:pla89}, the present polarimeter was
the first to measure the longitudinal polarization of a stored
polarized electron beam\cite{cbp:igo96}.

In section~\ref{sec:physics}, the physics of Compton scattering is
discussed with emphasis on the energy range of the AmPS ring and on
the technique for the measurement of longitudinal polarization. In
section~\ref{sec:design} the layout and the technical characteristics
of the polarimeter are described, and in section~\ref{sec:results} the
results of the performance tests of the polarimeter are presented.
Emphasis is put on the investigation of the systematic uncertainties
of the polarimeter. Conclusions and a summary are given in
section~\ref{sec:conclusions}.

\section{The Physics of Compton Scattering}
\label{sec:physics}

\begin{figure} 
\epsfig{figure=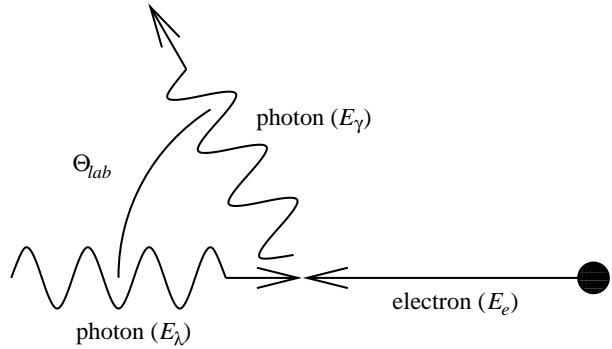,height=8cm,angle=-90}
\caption{The kinematics of Compton scattering in the lab frame.  The
initial photon and electron energies are expressed as $E_\lambda$ and
$E_e$, respectively, the energy of the scattered photon as $E_\gamma$, and
the angle between initial electron and final state photon as
$\theta_{lab}$.  The scattered electron is not shown.}
\label{cbpfig:kinLAB} 
\end{figure}

The kinematics of Compton scattering in the lab frame is shown in
fig.~\ref{cbpfig:kinLAB}.  The initial photon and electron energies
are expressed as $E_\lambda$ and $E_e$, respectively, while the
scattered photon energy is expressed as $E_\gamma$.  The scattered
electron is not shown.  The positive z-axis is defined to be the
direction of the incident electron.  Unlike the kinematics of
transversely polarized electrons, the reaction with longitudinally
polarized electrons is symmetric about the azimuthal angle $\varphi$,
and thus only the angle $\theta$, between the incident electron and
scattered photon, is shown.  In the case of high-energy electrons,
photons are backscattered in a narrow cone centered around the
direction of the initial electron. Typical values of the scattering
angle in the lab frame, corresponding to $90^\circ$ scattering in the
electron rest frame, are given in table~\ref{cbptab:theory}.

\begin{table}
\begin{tabular}{|c|c|c|c|c|c|}  \hline
 $E_e[\text{MeV}]$&  $E_{\gamma}^{max}[\text{MeV}]$& 
 $\theta_{crit}[\text{mrad}]$& $\sigma_{0}[\text{mbarn}]$&
 $\alpha_{3z}(E_{\gamma}^{max}) [10^{-3}]$& $\cos(\varphi_p)$\\ \hline
 500 & \,\,\,9.0  & 1.02 & 654 & 18.3   & 0.98 \\
 700 & 17.6               & 0.73 & 649 & 25.5   & 0.95 \\
 900 & 28.9               & 0.57 & 645 & 32.7   & 0.92 \\    \hline
\end{tabular}

\caption{Maximum energy of Compton backscattered photons
($E_{\gamma}^{max}$), scattering angle in the lab frame for photons
with 90$^\circ$
scattering in the electron rest frame ($\theta_{crit}$), total
unpolarized Compton cross section ($\sigma_{0}$), maximum value of
the spin correlation function ($\alpha_{3z}(E_{\gamma}^{max})$) and
reduction factor ($\cos(\varphi_p)$) for different electron beam
energies ($E_e$) which can be stored in the AmPS ring and for
$E_\lambda = 2.41$~eV (514~nm).}

\label{cbptab:theory}
\end{table}

\sloppypar The polarization of the electron is specified in Cartesian
coordinates by $\mathbf{P}=P_e\hat{\mathbf{P}}=(P_{x},P_{y},P_{z})$
and that of the incident photon by the Stokes vector
$\mathbf{S}=(S_{0},S_{1},S_{2},S_{3})$~\cite{cbp:fan49}.  The amount
of linearly polarized light is given by $\sqrt{S_{1}^2+S_{2}^2}$ and
that of circular polarization by $|S_{3}|$.  The sign of $S_{3}$
indicates the helicity of the polarization: $S_{3}>0$ for left-handed
helicity ($S_{3L}$) and $S_{3}<0$ for right-handed helicity
($S_{3R}$).  For normalization of the Stokes vector $S_0$ is taken to
be unity.
 
\begin{figure}
\epsfig{figure=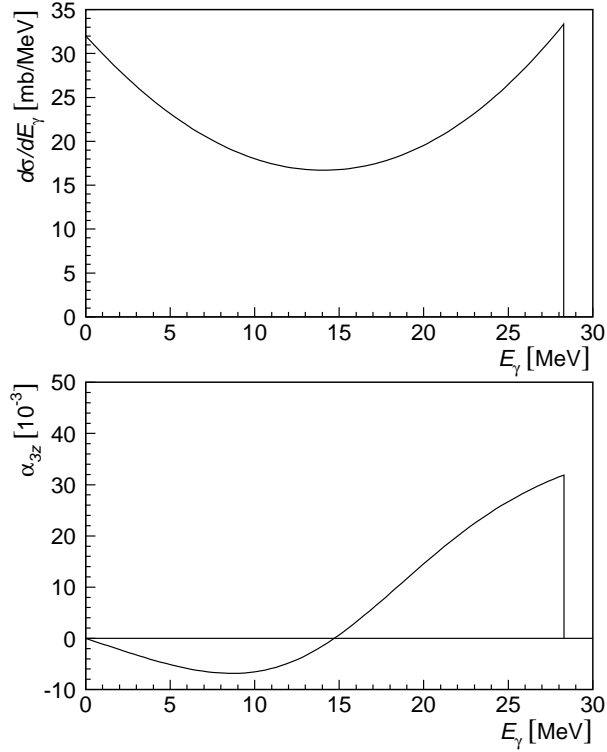,width=8cm}
\caption{Cross section and spin correlation function for Compton
  scattering, with $E_\lambda=2.41$~eV and
  $E_e=900$~MeV as function of the energy of the backscattered photon
  ($E_\gamma$).}
\label{cbpfig:cross}
\end{figure}

The cross section for Compton scattering of circularly polarized
photons by longitudinally polarized electrons can be written as:
\begin{equation}
\frac{d\sigma}{dE_{\gamma}}=
\frac{d\sigma_{0}}{dE_{\gamma}}[1+S_{3}P_{z}\alpha_{3z}(E_{\gamma})]
\label{eq:sigmapol}
\end{equation}
where $\alpha_{3z}(E_{\gamma})$ is the circular-longitudinal spin
correlation function and $\frac{d\sigma_{0}}{dE_{\gamma}}$ is the
energy-differential cross section for unpolarized electrons and
photons.  This cross section and the spin correlation function are
shown in fig.~\ref{cbpfig:cross}. Exact formula's for these quantities
can be found e.g.\ in
refs.~\cite{cbp:iz80,cbp:fernow86,cbp:lip54,cbp:tol56}.

Equation~\ref{eq:sigmapol} shows that an asymmetry measurement can be
performed by switching the sign of $S_3$ which gives an asymmetry
proportional with $P_z$. 
For a given $E_{\lambda}$ and $E_{e}$ this asymmetry can be written as,
\begin{equation}
  A(E_{\gamma})=\frac{\frac{d\sigma}{dE_{\gamma}}_L -
    \frac{d\sigma}{dE_{\gamma}}_R}{\frac{d\sigma}{dE_{\gamma}}_L +
    \frac{d\sigma}{dE_{\gamma}}_R}
  ={\Delta}S_3P_{z}\alpha_{3z}(E_{\gamma})
\label{eq:genasym}
\end{equation}
where $\frac{d\sigma}{dE_{\gamma}}_L$
($\frac{d\sigma}{dE_{\gamma}}_R$) is the polarized cross section with
incident left-handed (right-handed) helicity, and
${\Delta}S_3=\half(S_{L3}-S_{R3})$.
 
The longitudinal polarization of the electron beam, $P_{z}$, can be
determined by taking the quantity $P_{z}$ as a free parameter
while fitting the measured asymmetry with eq.~\ref{eq:genasym}.  
The photon polarization term, ${\Delta}S_3$, needs to be measured 
independently.

Typical values of the total unpolarized Compton cross section, the
maximum energy for backscattered photons, and the maximum value for the
spin correlation function are given in table~\ref{cbptab:theory} for
various electron beam energies that can be stored in the AmPS ring.

\section{The NIKHEF Compton Polarimeter}
\label{sec:design}

\begin{figure}
\epsfig{figure=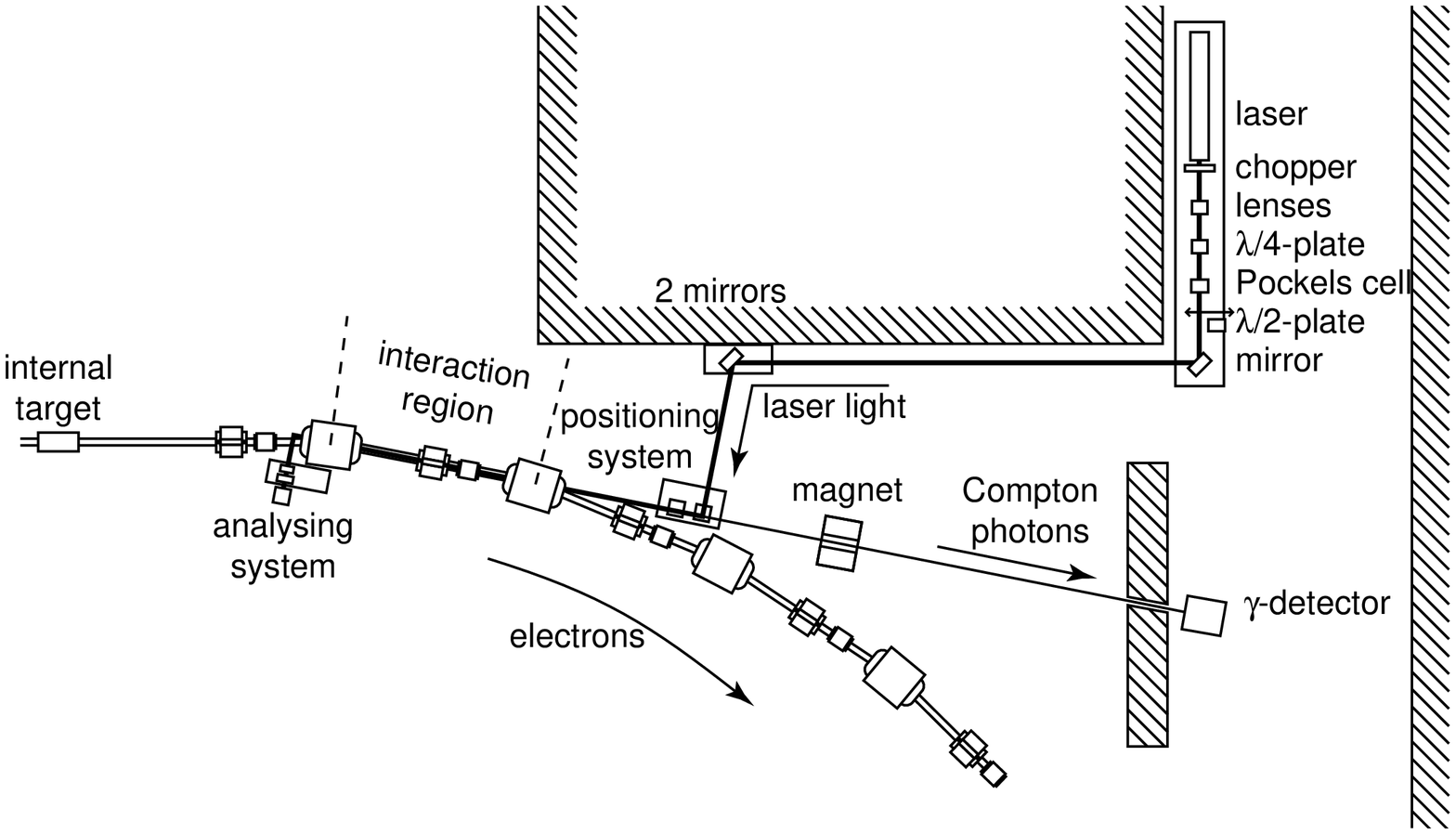,width=\textwidth}
\caption{The layout of the polarimeter setup showing the path of laser light
  and the detector for the Compton backscattered photons.}
\label{cbpfig:closeup}
\end{figure}

A schematic layout of the Compton polarimeter is shown in
fig.~\ref{cbpfig:closeup}.  The polarimeter consists of the laser
system with its associated optical system and a detector for the
detection of the backscattered photons\cite{cbp:vod96}.  Laser photons
interact with stored electrons in the straight section ($\sim 3$~m
long) after the first dipole (bending angle $11.25^\circ$) and before
the second dipole after the Internal Target Facility (ITF).  The
distance of the Interaction Region (IR) from the internal target is
4.5~m.  

The Siberian snake preserves the longitudinal spin direction at the
ITF, but the orientation of the spin precesses in the ring,
according to\cite{cbp:mon84}
\begin{eqnarray}
  \varphi_p&=&\frac{E_e\text{[MeV]}}{440.65} \varphi _e \\
  P_z &=&P_e \cos\varphi_p
\end{eqnarray}
where $\varphi_e$ is the bending angle of the electrons and
$\varphi_p$ the precession angle of the spin.  The polarimeter is only
sensitive to the longitudinal component of the electron spin (see
eq.~\ref{eq:genasym}).  Therefore, the experimental asymmetry of the
polarimeter reduces with $\cos(\varphi_p)$.  To minimize this
reduction factor, the polarimeter has to be located close to the ITF.
If for example the North or South straight (see
fig.~\ref{cbpfig:amps}) would be used for the polarimeter (90$^\circ$
rotation from the ITF), the spin would be exactly transverse at
$E_e$~=~440~MeV, and the reduction factor would be zero.  On the other
hand, because of the internal target at the ITF, the residual gas
pressure in the 32~m long west straight can be orders of magnitude
higher than in any other place in the ring, resulting in a large
bremsstrahlung background.  The section between the first and second
dipole after the ITF has been chosen as a compromise between spin
precession and bremsstrahlung background.

The backscattered photons leave the IR travelling in the same
direction as the electrons of the beam and are separated from them by
the second dipole.  The $\gamma$-photons leave the vacuum system via
a vacuum window and pass through a mirror of the optical system, a
sweeping magnet, and a hole in a concrete shielding wall to the gamma
detector.

\subsection{The laser and the associated optical system}
 
An Innova 100-25 Ar-ion laser system\footnote{Coherent Benelux,
  Argonstraat 136, 2718 SP Zoetermeer, The Netherlands} is used which
can provide a 10~W continuous beam.  The wavelength of the laser light
is $\lambda$~=~514~nm, the divergence 0.4~mrad (full angle), and the
diameter 1.8~mm.  The initially linearly polarized light is converted
to circularly polarized light with an anti-reflection coated
$\lambda/4$-plate.  Left- and right-handed light is obtained by
switching the high voltage on a Inrad 214-090 Pockels
cell\footnote{Inrad, 181 Legrand Avenue, Northvale, NJ 07647, USA},
positioned after the $\lambda/4$-plate, see fig.~\ref{cbpfig:closeup}.
The Pockels cell switches between zero-wave (if high voltage is off)
and half-wave (if high voltage is on) retardation.  The polarization
of the laser light is better than 99.8\% directly after the Pockels
cell.  We have observed a small degradation along the beam path, but
the degree of polarization is well above 95\% at the IR if the system
is tuned carefully.  After the first tests with a polarized electron
beam (see section~\ref{sec:results}), we have added a $\lambda/2$-plate
to our system, which reverses the sign of the laser polarization while
keeping all steering signals and the high voltage on the Pockels cell
constant.  The $\lambda/2$-plate can be moved into the optical path by
means of a compressed air driven translation stage and is located
directly after the Pockels cell.  It has been used to investigate
false asymmetries (see section~\ref{sec:results}).
 
To facilitate the transport of the laser beam and in order to prevent
the Pockels cell from damage by high intensity light, a Galilean expander
consisting of two lenses is placed in front of the $\lambda/4$-plate.
The focal lengths of the lenses are -2.5~cm and 8.0~cm with almost
coinciding focal points.  This results in an expansion of the laser
beam by a factor 3.2 and a reduction of the divergence by the same
factor.  The focal point of the expander is placed in the IR.  The
lenses are diffraction-limited and have anti-reflection coatings
optimized for 514~nm.
 
In order to protect both the Ar-ion laser and the Pockels cell from
radiation damage they are installed in a niche in the NW-curve of
AmPS.  The laser light is guided to the IR, over a total path of
approximately 12~m, by a system of six dielectric mirrors.  All
mirrors are optimized for 90$^\circ$ scattering at 514~nm.  The first
mirror reflects the beam from the niche in the direction of ITF.  A
set of two mirrors reflects the beam into the beam positioning system.
Two mirrors are used at this location in order to account for the
11.25$^\circ$ bend caused by the first dipole after ITF, while
maintaining 90$^\circ$ reflections on all mirrors.

\begin{figure}
   \epsfig{figure=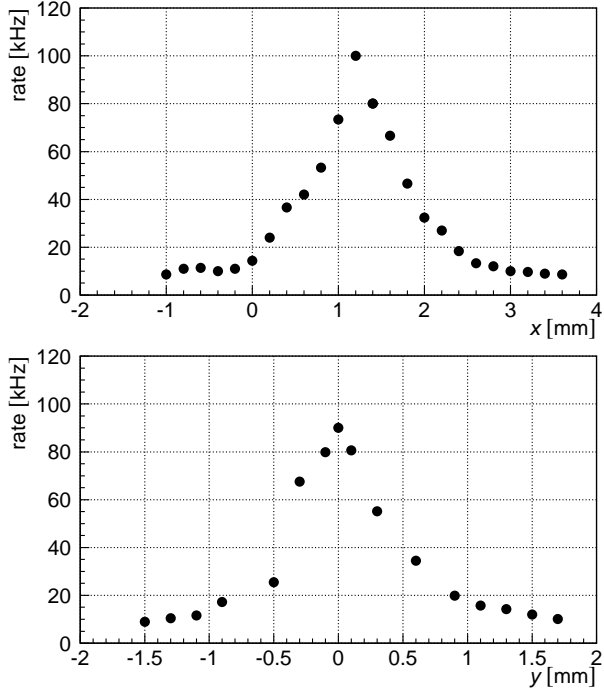,width=8cm}
   \caption{The rate of backscattered photons plotted versus laser
   beam position.}
   \label{cbpfig:xypos}
\end{figure}

The positioning system consists of two mirrors connected to two
translation stages and a gimbal mount.  Both translation stages and
the gimbal mount can be controlled with DC-motors.  The positioning
system gives full control over both angle and position of the laser
beam in all directions and is used to optimize the overlap of the
electron and laser beam.  Initial alignment was done manually,
centering it on the entrance and exit vacuum windows of the IR.  The
final alignment of the laser beam is performed with the DC-motors,
while electrons are circulating in AmPS.  The laser is aligned by
optimizing the rate of backscattered photons, see
fig.~\ref{cbpfig:xypos}.  A sixth mirror is used to reflect the beam
into the IR.

The laser beam enters and leaves the IR via Kodial ND-40 CF-F vacuum
windows\footnote{Balzers-Pfeiffer GmbH, Postfach 1280, D-35608 Asslar,
  Germany} (diameter 25~mm).  A system of two mirrors reflects the
beam into a beam analysis system after the second vacuum window.  It
consists of a power meter, a linear polarizer, and $\lambda/2$-plate.
The polarizer and $\lambda/2$-plate are mounted on a motorized
translation stage and can be taken out of the beam path to measure the
total transported laser power.  When the $\lambda/2$-plate and
polarizer have been inserted into the beam, the $\lambda/2$-plate can be
rotated with a DC-motor and the laser polarization can be determined
by measuring  the power as a function of the orientation of
the $\lambda/2$-plate. Laser beam polarization measurements done with
the beam analysis system are only used as a monitor of the system. The
polarization used in the extraction of the electron polarization from
an asymmetry measurement is measured manually immediately before and
after the vacuum windows of the IR. This ensures the absence of
systematic errors introduced by the two mirrors between the second
vacuum window and the beam analysis system.
 
A chopper mounted immediately after the laser is used to block the
laser light for $1/3$ of the time to measure the background.  The
chopper, operating at a frequency of 75~Hz, is also used to generate
the driving signal for the Pockels cell.  This ensures that the
background measurement and the flipping of the laser polarization have
exactly the same frequency.
 
The small fraction of the laser light transmitted through the mirrors
is utilized to monitor the position of the laser beam using camera's
which are located directly before and after the IR.
 
\subsection{The gamma detector}
 
The gamma detector of the polarimeter, situated $\sim$12~m downstream
of the IR, must be capable of detecting $\gamma$-rays with an energy
of up to 30~MeV (see table~\ref{cbptab:theory}), handling rates up to
1~MHz, and be radiation resistant.  Total absorption shower counters
made of dense inorganic scintillating crystals were chosen.

For the commissioning experiments with an unpolarized stored electron
beam a cylindrical BGO scintillator crystal (102~mm diameter and
100~mm long), optically coupled to a Hamamatsu R1250
photomultiplier\footnote{Hamamatsu Photonics Deutschland GmbH,
  Arzbergerstr.  10, D-8036 Herrsching am Ammersee, Germany} was used
to test and tune our laser positioning system.  A disadvantage of BGO
crystal is the long decay time (300~ns) of the scintillation light
which limits the maximum rate.
 
The BGO crystal was later replaced by a rectangular ($24 \times 10
\times 10$~cm$^3$) pure CsI crystal.  This crystal was optically
coupled to a XP4312/B photomultiplier\footnote{Philips Photonics, BP
  520, F-19106 Brive, France}.  Due to the small decay time of the
pure CsI crystal (35 and 6~ns), an active base on the photomultiplier
tube, and dedicated electronics, the total setup is able to handle
rates up to 1~MHz.  A 10~cm thick lead cylinder surrounds the
scintillator crystal for shielding purposes.  A plastic scintillator
(NE102) placed in the front of the gamma detector can be used to veto
charged particles.
 
\subsection{The Control and the Data Acquisition System}
 
\begin{figure}
  \epsfig{figure=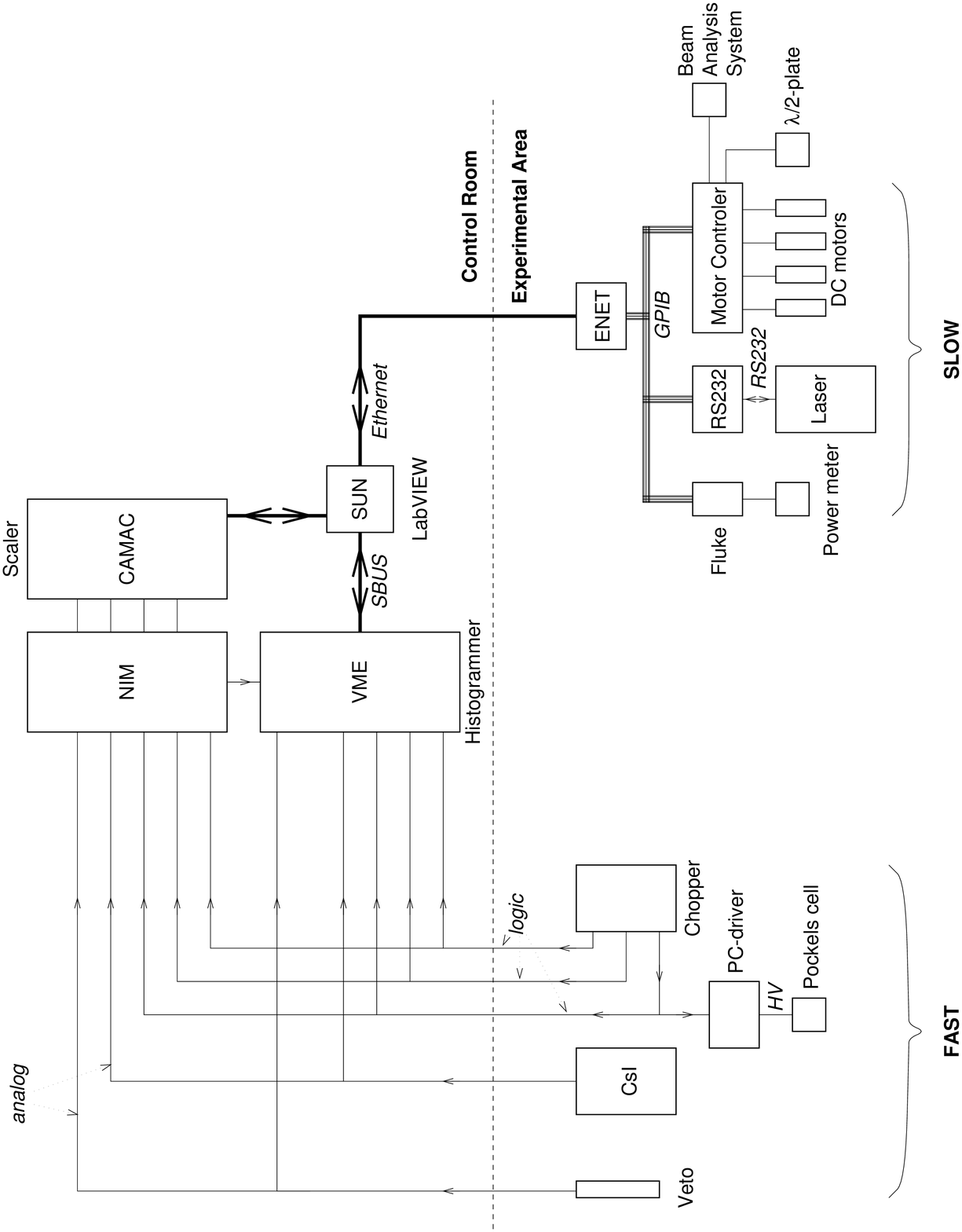,height=\textwidth,angle=-90}
  \caption{Schematic of the DAQ- and control-system of the Compton 
    polarimeter}
  \label{cbpfig:daq}
\end{figure}

The optical system is controlled with a computer code
developed with LabVIEW\footnote{National Instruments, 6504 Bridge
  Point Parkway, Austin, TX 78730-5039, USA} on a SUN-workstation, see
fig.~\ref{cbpfig:daq}. The communication between the workstation and the
optical system is realised via GPIB (IEEE-488.2), which is
connected to the workstation via Ethernet.  The four DC-motors used
for the steering of the laser beam are controlled by a PI804 motor
controller\footnote{Physik Instrumente GmbH, Polytec-Platz 5-7,
  D-76333 Waldbronn, Germany}.  A resolution of 2~$\mu$m and
60~$\mu$rad can be obtained with the PI804 and the DC-motors. Control
signals for the movement of the $\lambda/2$-plate after the Pockels
cell, and of the $\lambda/2$-plate and translation stage of the beam
analysis system are also generated in the PI804.
 
The data acquisition is done with a single VME-module, designed and
constructed at NIKHEF.  The module accepts the analog output of the
photomultiplier connected to the gamma detector and digitises this
signal using an 8-bit flash ADC.  Furthermore, it accepts logic
signals representing the state of the chopper and the Pockels cell.
Based on those signals, the module constructs four separate energy
spectra, for laser blocked or not and for left and right circularly
polarized light.  The maximum rate the module can handle is 2.5~MHz.
The energy spectra are read out typically every 30~s. The VME-module
can generate its own trigger for the ADC or it can accept an external
trigger. When the external trigger is used, more sophisticated
triggering systems can be made, while the internal trigger makes any
external logic modules superfluous, resulting in a very compact
system. An extra advantage of the external trigger circuit is the
possibility to connect the trigger not only to the VME-module, but
also to a scaler. The scaler information can then be used for the
optimization of the electron and laser beam overlap and to determine
dead-time corrections for all energy spectra. Both modes of operation
have been used.

Determining the electron polarization from the energy spectra is done
via eq.~\ref{eq:genasym}.  All energy spectra are normalized to their
integrated luminosities, taking into account dead-time effects and
measuring times. Background spectra are subtracted, after which the
experimental asymmetry ($A_{exp}$) is constructed, via:
\begin{equation}
  \label{eq:expasym}
  A_{exp}(E_\gamma) = \frac{n_L - n_R}{n_L + n_R} = \Delta S_3 P_e \cos
  \varphi_p \alpha_{3z}^{exp}
\end{equation}
where $n_L$ ($n_R$) are the energy spectra for left (right) polarized
laser light, after normalisation and background subtraction.  The
experimental spin correlation function $\alpha_{3z}^{exp}$ is obtained
with Monte Carlo simulations, performed with the computer code
GEANT\cite{GEANT}, taking into account the electron and laser beam
phase space along the IR and the characteristics of the gamma
detector. $\Delta S_3$ is measured separately and $\cos \varphi_p$ is
calculated from the electron beam energy. $P_e$ is extracted from
$A_{exp}$ via a fit with eq.~\ref{eq:expasym} with $P_e$ as a free
parameter. The results of the Monte Carlo simulations have also been
used for the energy calibration of the gamma detector.

\section{The performance of the polarimeter}
\label{sec:results}

The polarimeter has been tested with polarized electrons at a low beam
current of appr. 5~mA. These preliminary tests were the first direct
measurements of longitudinal polarization in an electron storage ring.
They are described in section~\ref{sec:prelimtests}. The accuracy of
the electron polarization enters directly in the accuracy of all
experiments performed with the polarized electron beam.  Therefore, it
is of the utmost importance that the statistical and systematic
uncertainty in $P_e$ is small and well known.  Statistical accuracy of
1\% can be obtained in 1500~s and will not influence the accuracy of
the internal target experiments significantly.  We have performed
extensive investigations of the systematic uncertainty of the
polarimeter, in order to minimize and understand these errors.  The
results of these investigations are presented in
section~\ref{sec:systematics}.  Finally, the polarimeter has been used
to monitor the electron polarization during the first internal target
experiment at NIKHEF that made use of the polarized beam. This is
presented in the last paragraph of this section.

\subsection {Preliminary tests}
\label{sec:prelimtests}

\begin{figure}
  \epsfig{figure=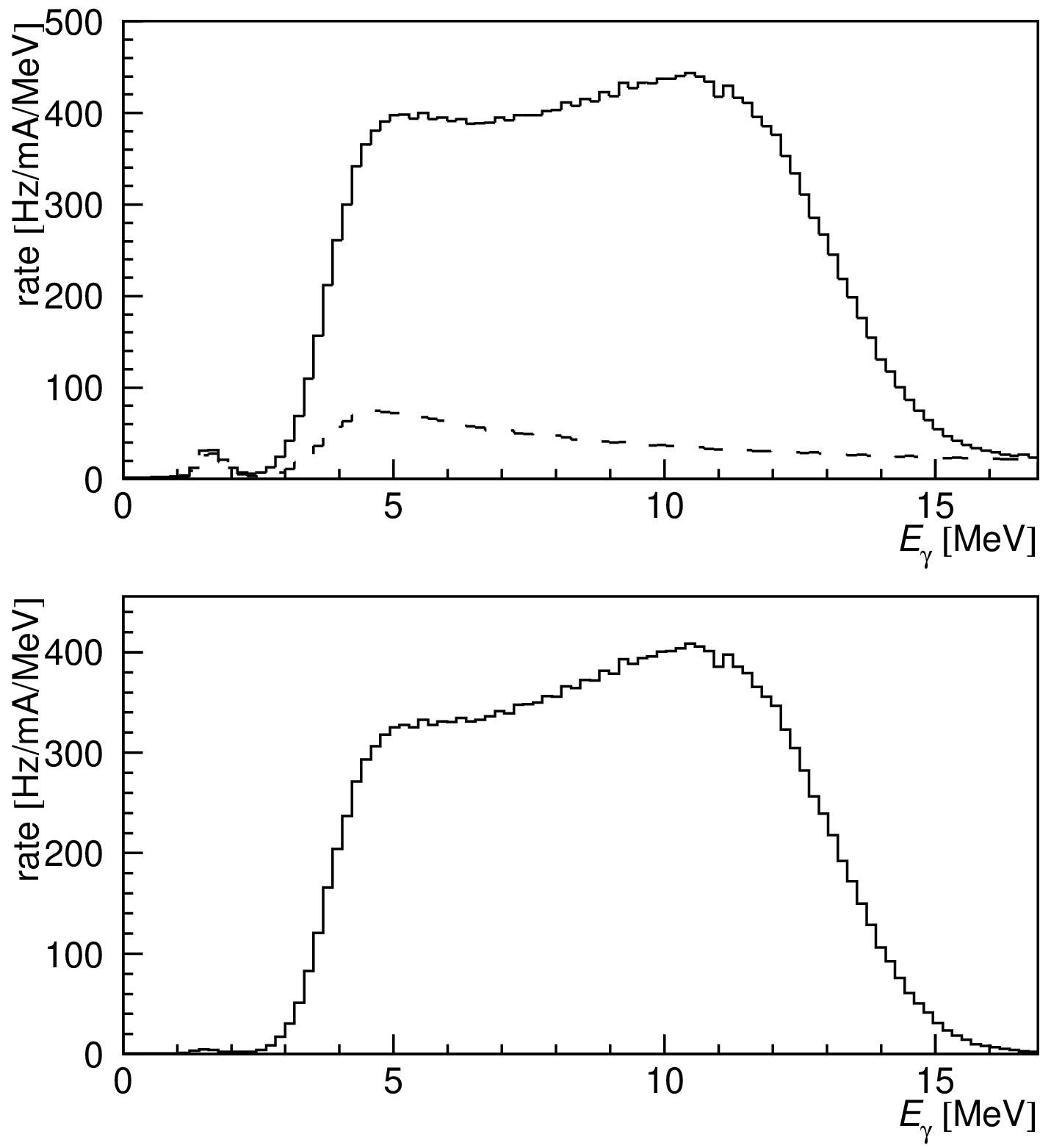,width=8cm}
\caption{Raw energy spectrum with laser on and laser off(top) and
  spectrum after subtraction of the background(bottom). All spectra
  are normalized to the beam current. The energy of the stored
  electron beam was 615~MeV.}
\label{cbpfig:raw}
\end{figure}

Preliminary tests with a polarized electron beam have been done with a
beam current of 2--5~mA and 3~W laser power, to avoid rate problems in
the gamma-detector and to minimize the contribution of background
events.  Because of the extremely low current in the accelerator and a
very poor injection efficiency, multiple injections were stacked to
obtain the beam current mentioned. The electron energy was 615~MeV,
resulting in backscattered photons with a maximum energy of
$E_{\gamma}^{max}=13.7$~MeV.  88\% of all detected photons originated
from Compton scattering, with a rate normalized to the beam current of
3.5~kHz/mA, see fig.~\ref{cbpfig:raw}.
 
The polarization of the electrons was measured with the Mott
polarimeter, located between PES and MEA, to be 0.41.  The orientation
of the electron spin at the injector was not optimized for maximum
polarization in AmPS, reducing the polarization in AmPS to
0.39\footnote{The difference in the polarization as measured with the
  Mott polarimeter cited in \cite{cbp:igo96} originates from a
  recalibration of the Mott polarimeter\cite{cbp:boris98}.}.
 
\begin{figure}
  \epsfig{figure=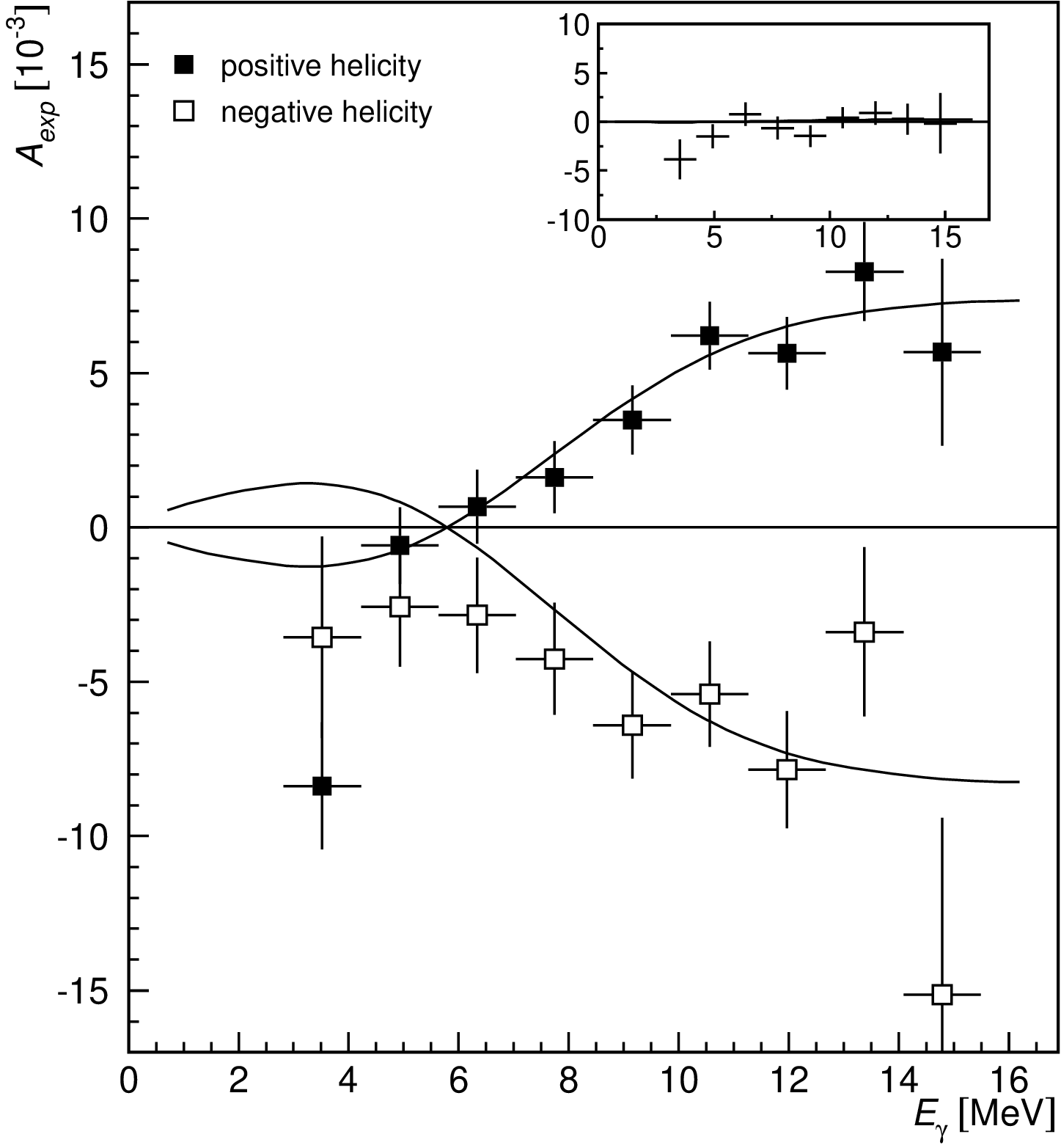,width=8cm}
\caption{Measured asymmetries as a function of photon energy ($E_{\gamma}$)
  for positive and negative electron helicity and for unpolarized
  electrons (inset).  The line is the best fit of 
  $P_{z}\Delta S_3 a_{3z}^{exp}$ to $A_{exp}$, as described in the text.}
\label{cbpfig:asym}
\end{figure}

During the experiment, an energy shift was observed between the energy
spectra for left- and right-handed polarized laser light, due to
ground currents (see next section).  This shift of 22~keV ($\approx
10^{-3}$ of the full scale) was corrected.  The asymmetry
$A_{exp}$ was measured and the electron polarization $P_e$ extracted,
see fig.~\ref{cbpfig:asym}.  In the extraction of the electron
polarization, we only used the values of $A_{exp}$ measured in the
energy range from 5 to 14~MeV.  The lower limit of the range is
determined by the energy threshold of the detector.  As the energy
shift described above occurs before the trigger is applied, the
asymmetry gets a systematic offset at the threshold (see the lowest
points in fig.~\ref{cbpfig:asym}).  The upper limit is chosen such
that the real to background ratio is better than 4:1, to avoid
sensitivity to the background subtraction.
 
We measured $P_e =0.38 \pm 0.04$ ($P_e=-0.42 \pm 0.06$) for electrons
with positive (negative) helicity. These numbers are in good agreement
with the result from the Mott polarimeter. We also determined
$A_{exp}$ for unpolarized electrons and found $P_e=0.01 \pm 0.04\%$,
indicating the absence of false asymmetries in our measurements.  This
first result obtained with the Compton polarimeter proves that it is
possible to accelerate polarized electrons and inject them in a
storage ring, even if stacking is required. It also shows that it is
possible to operate a Compton polarimeter to determine the absolute
degree of polarization of a longitudinally polarized stored electron
beam.

\subsection{Systematic checks}
\label{sec:systematics}

We have performed a series of measurements to investigate the
systematic errors of the polarimeter.  False asymmetries result in
errors on the determination of the electron polarization.  Possible
sources of false asymmetries are: a) inaccuracies in the ratio of the
integrated luminosities for the two laser polarization states; b)
inaccuracies in the determination of the background contribution, and;
c) any signal in phase with the asymmetry measurement.  The first type
of false asymmetry will give a energy-independent contribution, while
type~b will depend on $E_\gamma$.  Type~c can either be energy
dependent or independent, based on how the signal influences the
asymmetry measurement.

During these measurements, the storage ring could only be operated
with a 10\% partial snake\cite{cbp:ohm96}.  Therefore, it was
necessary to perform all measurements at an electron beam energy of
440~MeV, resulting in a maximum energy for the Compton photons of
7.0~MeV.  This beam energy is lower than the design specification of
the polarimeter (500--900~MeV), resulting in a poor energy resolution.
To reduce background at this rather low energy, we performed all
measurements with beam currents smaller than 15~mA, resulting in
Compton rates $\leq$~120~kHz (8.0~kHz/mA).  An advantage of the low
beam energy for our measurements is an enhanced sensitivity for false
asymmetries, because of the relatively small Compton asymmetry (see
table~\ref{cbptab:systematics}). The polarization measured with the
Mott polarimeter was 0.69. The settings of the Z-shaped spin
manipulator were optimized for maximum polarization in the AmPS
ring\cite{hjb98}.

\begin{table}
  \begin{tabular}{|l|c|c|} \hline
    source of systematic error       & $\Delta P_e$ \\ \hline
    $E_\gamma$ calibration           & 0.022 \\
    $P_{laser}$                      & 0.013 \\
    $\alpha_{3z} $ parametrisation   & 0.004 \\
    $E_e$                            & 0.003 \\
    Energy spectrum shift            & 0.001 \\
    Luminosity asymmetry             & 0.001 \\ \hline 
    Total                            & 0.027 \\ \hline
  \end{tabular}
  \caption{All sources of systematic errors contributing to the error on
    $P_e$. The errors are calculated for 440~MeV electrons with $P_e =
    0.60$. See the text for an explanation of the sources of the
    systematic errors.}
  \label{cbptab:systematics}
\end{table}

The measurements showed an energy-independent asymmetry of the order
of $0.5\cdot 10^{-3}$.  It disappeared if we disabled the switching of
the Pockels cell.  After we switched the Pockels cell the rate changed
of the order of 10\% in 10~s, and then stabilized.  If we moved the
laser beam by 0.1~mm simultaneously with the switching of the Pockels
cell, the equilibrium rate was the same for both states.  From this,
we concluded that the Pockels cell is heated by applying the high
voltage.  The effect is a small steering of the laser beam, in the
order of 10~$\mu$rad between the two equilibrium states, or 15~nrad
during normal operation, resulting in a false asymmetry of type~a.
The exact value of this false asymmetry changed over the time of days,
though the order of magnitude was constant.  It can be explained by
temperature drifts of the laser, because the size of the false
asymmetry depends strongly on the exact position of the laser beam
with respect to the electron beam.

False asymmetries from inaccuracies in the background contributions
are negligible, because of the good real to background ratio in the
energy range used for the polarization measurements.  Furthermore, the
background has to be related to the laser polarization in order to
introduce false asymmetries.  In the energy range used, no physics
process can contribute significantly to the asymmetry.  The background
subtraction could also introduce an error in the size of the
asymmetry, if the background subtraction was not performed accurately.
Also this effect is negligible, because we measured the background
simultaneously, the background is small compared to the real signal,
and we correct all energy spectra for dead-time effects.

The only signal in phase with the asymmetry measurements is the
driving signal of the Pockels cell and electronics.  We have observed
a false asymmetry of the order of $2\cdot 10^{-3}$, related to the
driving signal of the Pockels cell and electronics.  It is
proportional to the derivative of the energy spectrum.  This indicates
that the signal from the gamma detector is shifted by the driving
signal of the Pockels cell, before digitisation.  This can happen
either between the detector and the electronics, or inside the
VME-module.  The asymmetry disappears if we generate the driving
signals for the electronics separately, indicating that the shift of
the analog signal happens before the signal reaches the electronics.
This is confirmed by the fact that the false asymmetry also disappears
if the signal from the gamma detector is disconnected from the
electronics.  The shift of the analog signal results in a shift of the
energy spectrum of the order of $1\cdot 10^{-3}$ of the full energy
scale.

\begin{figure}
  \epsfig{figure=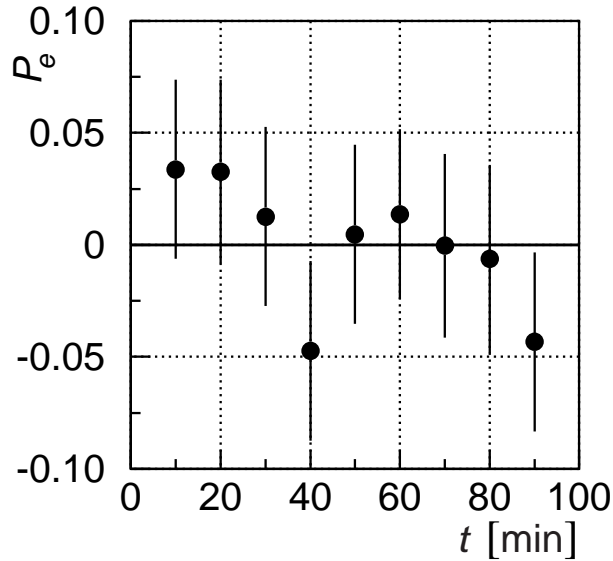,width=8cm}
\caption{Short-term stability of the polarimeter. Every data point
  represents a measurement of the polarization of unpolarized
  electrons. The time between two measurements is appr.\ 10 minutes.}
\label{cbpfig:reproshort}
\end{figure}

Both false asymmetries mentioned above can be corrected for, if they
are known during real polarization measurements.  Therefore, we have
performed all measurements in sets of six.  Three measurements were
done with different electron polarizations injected into the ring
(positive helicity, unpolarized and negative helicity).  These
measurements were repeated with the $\lambda/2$-plate inserted in the path
of the laser beam, which reverses the sign of the measured Compton
asymmetry by a change of sign of the laser polarization.  The
measurements with unpolarized electrons were used to determine the
false asymmetries.  This is only valid, if the false asymmetries do
not change on the time scale of the 2 times 3 measurements.  To check
this, one measurement of such a set was repeated nine times.  To
exclude sensitivity to variations in the polarization of the injected
electrons or spin life time, we choose to use unpolarized electrons.
The total measurement time was appr.\ 90 minutes., while a full set of
six measurements normally takes about 60 minutes.  The results are shown
in fig.~\ref{cbpfig:reproshort} and show good stability on this time
scale, indicating that we can use measurements with unpolarized
electrons to correct for false asymmetries.

The systematic uncertainty of the polarimeter is not only determined
by false asymmetries, but also by the analysis parameters.  The main
contribution comes from the energy calibration of the detector.
Smaller effects come from the uncertainty in the energy of the
electron beam, the laser polarization and the parametrisation of the
theoretical analysing power, folded with the parameters of the
detector, like the energy resolution.  The laser polarization measured
is the average polarization of the whole laser beam.  The laser beam
is larger than the electron beam and therefore only its central part
interacts with the electrons.  We have measured the laser polarization
in a range of transverse positions and did not observe any variation.

Table~\ref{cbptab:systematics} shows an overview of all sources of
systematic errors for electrons of 440~MeV with a polarization of
0.60.  The systematic error will decrease for higher beam energy,
because the Compton asymmetry increases.

\subsection{Polarization monitoring}
During the first data run of experiment 94-05~\cite{cbp:prop9405}, the
polarimeter has been used to monitor the polarization of the electrons
stored in the AmPS ring. This data run followed immediately after the
measurements described in the previous section. These measurements
were also performed with a partial snake at a beam energy of 440~MeV.
The polarization could not be measured simultaneously with the
experimental data taking, because the background rate due to $^3$He
gas leaking into the IR from the internal target was too high.
Therefore, the electron polarization was measured once a day when no
gas was fed into the internal target.

\begin{figure}
  \epsfig{figure=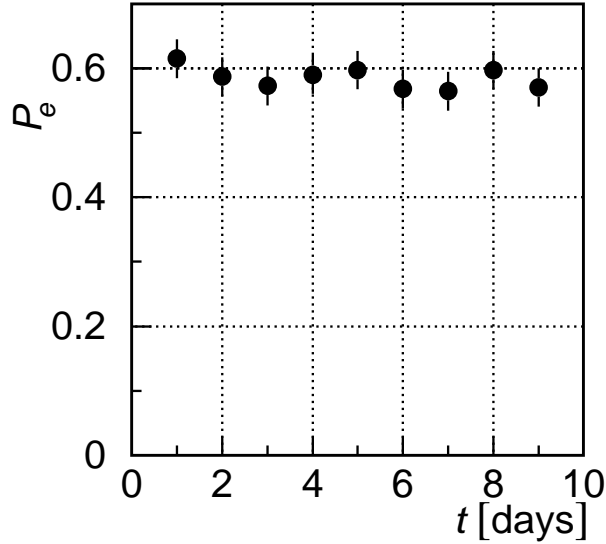,width=8cm}
\caption{Long-term stability of the polarimeter.  Every data point
  represents a complete set of six polarization measurements as
  described in the text.  The interval between the measurements is
  typically one day.}
\label{cbpfig:reprolong}
\end{figure}

The long-term stability was investigated from these measurements.
They are sensitive not only to variations of the polarimeter, but also
to any other time-dependent effect such as a degradation of the
photo-cathode used at PES. No trend is observed in the polarization of
the electrons (see fig.~\ref{cbpfig:reprolong}), indicating a good
long-term stability for all components, including the polarimeter.

The average polarization for all measurements and for negative and
positive electron helicity determined with the $\lambda/2$-plate in
(out) the laser beam was measured to be $0.615 \pm 0.009 \pm 0.027$
($0.595 \pm 0.009 \pm 0.027$). This is less than the value measured
with the Mott polarimeter (0.69). No polarization loss was observed
during the first measurement at 615~MeV (see
section~\ref{sec:prelimtests}). The difference between the two
measurements were the beam energy, the settings of the snake, and the
orientation of the spin in the accelerator.

The beam energy has no direct effect on the depolarization in the
accelerator.  Depolarization can occur in the focussing solenoids in
the first sections of the accelerator. Small energy differences of the
electrons can cause differences in the spin precession in the lenses,
which will result in loss of coherence of the transverse component of
the electron spin. At 615~MeV the spin was longitudinal in the
accelerator and thus was not sensitive for this effect. The electron
spin in the measurements at 440~MeV was completely transverse and so
had maximum sensitivity for depolarization due to the lenses.

The difference might also be explained by polarization losses during
injection, due to the partial snake. A full snake ensures that the
spin tune is 0.5, so that all possible spin resonances are far away.
The spin tune with the partial snake was 0.05. If this was
close to a depolarizing resonance, it is possible that during
injection the spin life time is shorter due to larger synchrotron
oscillations and so cause a loss of polarization during the damping of
the beam. We have measured the spin life time of the damped
beam, to be well over 3600~s which can not explain the losses we
have observed.

\section{Summary - Conclusions}
\label{sec:conclusions}
 
We have successfully designed, constructed, and operated a Compton
polarimeter to measure the longitudinal polarization of a stored
electron beam. We have measured the polarization of a stored beam at
beam energies of 440~MeV and 615~MeV. The absolute systematic
uncertainty has been determined to be 0.027 for electrons at 440~MeV
with a polarization of 0.60.  The systematic uncertainty will decrease
at higher beam energies.  It has been shown that the polarimeter can
be operated routinely and reliably during the first experiment at the
internal target facility that made use of polarized electrons.  Extra
pumping capacity will be installed in the future to make polarization
measurements possible during operation of an internal target.

\ack
 
We would like to thank C.  L.  Morris (LANL) for the supply of the CsI
crystal and O. Hausser (DESY) for advise on the optical system.  We
would also like to thank the NIKHEF Electronic workshop for the design
and construction of the data acquisition module. We like to thank the
accelerator group for their assistance with the operation of the
facility, the PES group for the polarized beam and the 94-05
collaboration for the time and manpower they made available to perform
the experiments discussed. This work was supported in part by the
Stichting voor Fundamenteel Onderzoek der Materie (FOM), which is
financially supported by the Nederlandse Organisatie voor
Wetenschappelijk Onderzoek (NWO), and by HCM Grant Nrs.  ERBCHBICT-930606 and
ERB4001GT931472.

\clearpage %
\onecolumn %

\bibliography{polarimeter} %

\begin{thebibliography}{10}

\bibitem{cbp:bar93}
D.~P. Barber {\em et~al.},
\newblock Nucl. Instr. Meth. Phys. Res. {\bf A329} (1993) 79.

\bibitem{cbp:knu91}
L.~Knudsen {\em et~al.},
\newblock Phys. Lett. {\bf B270} (1991) 97.

\bibitem{cbp:bol96}
Y.~B. Bolkhovityanov {\em et~al.},
\newblock The polarized electon source at {NIKHEF},
\newblock in {\em Proc. of the $12^{th}$ International Symposium on High Energy
  Spin Physics}, edited by C.~W. de~Jager {\em et~al.}, pages 730--732, World
  Scientific, 1996.

\bibitem{cbp:der73}
Ya.~S. Derbenev and A.~M. Kondratenko,
\newblock Sov. Phys.-JETP {\bf 37} (1973) 968.

\bibitem{cbp:der75}
Ya.~S. Derbenev and A.~M. Kondratenko,
\newblock Sov. Phys.-Dokl. {\bf 20} (1975) 830.

\bibitem{jps97}
C.~W. de~Jager, V.~Ptitsin and Yu.~M. Shatunov,
\newblock Radiative electron polarization in the {AmPS} storage ring,
\newblock in {\em Proc. of the $12^{th}$ International Symposium on High-Energy
  Spin Physics}, edited by C.~W. de~Jager {\em et~al.}, pages 555--557, World
  Scientific, 1996.

\bibitem{cbp:bar94}
D.~P. Barber {\em et~al.},
\newblock Nucl. Instr. Meth. Phys. Res. {\bf A338} (1994) 166.

\bibitem{cbp:pla89}
M.~Placidi and R.~Rossmanith,
\newblock Nucl. Instr. Meth. Phys. Res. {\bf A274} (1989) 79.

\bibitem{cbp:igo96}
I.~Passchier {\em et~al.},
\newblock A {C}ompton backscattering polarimeter for electron beams below 1
  {GeV},
\newblock in {\em Proc. of the $12^{th}$ International Symposium on High-Energy
  Spin Physics}, edited by C.~W. de~Jager {\em et~al.}, pages 807--809, World
  Scientific, 1996.

\bibitem{cbp:fan49}
U.~Fano,
\newblock Optical Society of America {\bf 39} (1949) 859.

\bibitem{cbp:iz80}
C.~Itzykson and J.~Zuber,
\newblock {\em Quantum Field Theory},
\newblock McGraw-Hill Inc., 1980.

\bibitem{cbp:fernow86}
R.~C. Fernow,
\newblock {\em Introduction to Experimental Particle Physics},
\newblock Cambridge University Press, 1986.

\bibitem{cbp:lip54}
F.~W. Lipps and H.~A. Tolhoek,
\newblock Physica {\bf 20} (1954) 85.

\bibitem{cbp:tol56}
H.~A. Tolhoek,
\newblock Rev. Mod. Phys. {\bf 28} (1956) 277.

\bibitem{cbp:vod96}
N.~P. Vodinas {\em et~al.},
\newblock Electron beam polarimeter,
\newblock in {\em International Workshop on Polarized Beams and Polairzed
  Targets}, edited by Hans~Paetz gen. Schieck and Lutz Sydow, pages 328--332,
  World Scientific, 1996.

\bibitem{cbp:mon84}
Bryan~W. Montague,
\newblock Phys. Rep. {\bf 113} (1984) 1.

\bibitem{GEANT}
{GEANT team},
\newblock {\em {GEANT} - Detector Description and Simulation Tool},
\newblock CERN Geneva, Switzerland,
\newblock version 3.21.

\bibitem{cbp:boris98}
B.~L. Militsyn,
\newblock private communication.

\bibitem{cbp:ohm96}
C.~Ohmori {\em et~al.},
\newblock Phys. Rev. Lett. {\bf 76} (1996).

\bibitem{hjb98}
H.~J. Bulten {\em et~al.},
\newblock Phys. Rev. Lett.  (1998),
\newblock to be submitted.

\bibitem{cbp:prop9405}
R.~Alarcon {\em et~al.},
\newblock {NIKHEF} proposal 94-05, 1994,
\newblock spokesman: J. F. J. van den Brand.

\end{thebibliography}

\end{document}